\documentclass[aps,prb,twocolumn]{revtex4}%
\usepackage{amsfonts}
\usepackage{eurosym}
\usepackage{amsmath}
\usepackage{txfonts}
\usepackage[scaled]{helvet}
\usepackage{color}
\usepackage[breaklinks=true,colorlinks,citecolor=blue,linkcolor=blue,urlcolor=blue]%
{hyperref}
\usepackage[pdftex]{graphicx}
\usepackage{float}
\usepackage{color}
\usepackage{rotating}
\usepackage{graphicx}
\usepackage{natbib}
\usepackage{dcolumn}
\usepackage{bm}
\usepackage{hyperref}
\usepackage[mathlines]{lineno}
\usepackage[caption=false]{subfig}
\usepackage{epstopdf}
\usepackage{amssymb}%
\providecommand{\U}[1]{\protect\rule{.1in}{.1in}}
\setcitestyle{numbers,square}

\begin{document}
\title{Magnon dispersion in bilayers of two-dimensional ferromagnets}
\author{Lara C. Ortmanns$^{1,2,3}$, Gerrit E. W. Bauer$^{1,4}$, and Yaroslav M.
Blanter$^{1}$}
\affiliation{$^{1}$Kavli Institute of Nanoscience, Delft University of Technology,
Lorentzweg 1, 2628 CJ Delft, The Netherlands }
\affiliation{$^{2}$ Institute of the Theory of Statistical Physics, RWTH Aachen, 52056
Aachen, Germany}
\affiliation{$^{3}$ Department of Microtechnology and Nanoscience (MC2), Chalmers
University of Technology, SE-41298 G\"{o}teborg, Sweden}
\affiliation{$^{4}$ WPI-AIMR and Institute for Materials Research and CSRN, Tohoku
University, Sendai 980-8577, Japan}

\begin{abstract}
\noindent We determine magnon spectra of an atomic bilayer magnet with
ferromagnetic intra- and both ferro- and antiferromagnetic interlayer
coupling. Analytic expressions for the full magnon band of the latter case
reveal that both exchange interactions govern the fundamental magnon gap. The
inter and intralayer magnetic ordering are not independent: a stronger
ferromagnetic intralayer coupling effectively strengthens the antiferromagnetic
interlayer coupling as we see from comparison of two bilayer systems. The trivial topology of these exchange-anisotropy spin
models without spin-orbit interaction excludes a magnon thermal Hall effect.

\end{abstract}
\maketitle

\section{Introduction}

\label{section:intro}
Two-dimensional van der Waals magnets (2DvdWM) \cite{Park} are a unique
platform to study magnetism in 2+$\varepsilon$ dimensions
\cite{Soriano2020Sep,2dvdwreview,lit:rev_morpurgo}. Two-dimensional order is
associated with strong intrinsic thermal fluctuations
\cite{Mermin-Wagner,2dvdwreview} and characteristic quantum phases
\cite{2dvdwreview}, offering a new test bed for competing interactions, such
as Heisenberg and anisotropic exchange \cite{fernandez-rossier} with different
range, Dzyaloshinskii-Moriya (DM) \cite{DMI1,DMI2} and other spin-orbit
couplings \cite{Kitaev1,Kitaev2}, and magnetodipolar interactions
\cite{Johnston,Bruno} in a rich variety of elements and crystal structures.
The parameters of many properties are highly tunable by electric gating
\cite{deng,huang} or by strain \cite{Guinea,Ma}. Of particular interest is the
control of the magnetic anisotropy that modulates the spin fluctuations and
allows to study cross-overs between different types of spin Hamiltonians
\cite{2dvdwreview}. 2DvdWM can be stacked with themselves or other materials
into multilayers \cite{mcguire, soriano, lit:nanolett_cri3, jiang} or
structured into nanodevices and directly accessed by scanning probe microscopy
or other surface sensitive experimental techniques \cite{lit:ref39}.

In this young field, many basic questions are still open. Only recently the
magnon energy dispersion has been calculated, which is essential for
understanding the spin dynamics and transport \cite{Owerre}.
For compounds with a hexagonal lattice such as CrI$_{3}$ and CrBr$_{3}$
\cite{crystals} as considered here we may expect a magnon dispersion relation
similar to that of the $\pi$-electron bands of graphene --- a minimum at $k=0$
and two degenerate Dirac points per unit cell at an intermediate energy. This
was confirmed by an analytic expression for a 2DvdWM with ferromagnetic (FM)
exchange interactions \cite{lit:owerre,Pershoguba,Aguilera}. However, bilayers
with FM intra- and inter-layer exchange interaction show characteristic
differences with bilayer graphene in terms of the degeneracy and dispersion
close to the Dirac points \cite{Owerre}. To date, the magnon dispersion for
bilayers with antiferromagnetic (AFM) coupling has to the best of our
knowledge been computed only numerically \cite{Owerre,Zhai}.

Here we extend previous theories by including a more general form of the
perpendicular plane magnetic anisotropy.
For the bilayer with FM intra- and AFM inter-layer exchange, we report
\textit{analytical} results for the full spectrum by a method introduced by Colpa
\cite{lit:colpa}. We analyze the interplay of FM intra- and AFM interlayer
couplings as reflected in the fundamental gap and total energy. The analytic
solutions facilitate access to non-trivial topological properties such as the
magnon Hall effect. For the class of perpendicular-plane anisotropy models
without magnetization texture or spin-orbit interaction the topology is
trivial, however.

The manuscript is organized as follows: In Sec.\ref{section:model}, we define
the most general spin Hamiltonian of 2DvdWM. In Sec.\ref{section:ferro}, we
review results on magnon spectra of an FM monolayer with different types of
anisotropy and a bilayer with FM intra-and interlayer coupling. In
Sec.\ref{section:antiferro} we present our main results, i.e., an analytic
derivation of the dispersion for a bilayer with FM intra-and AFM interlayer
exchange coupling. We consider first isotropic exchange coupling for different
spin configurations and subsequently include perpendicular spin anisotropy. We
analyze the effect of the magnetic order on the fundamental gap as well as
total energy. Finally we compute the magnon Chern numbers of the energy bands.
Sec.\ref{section:conclusions} summarizes our conclusions and gives an outlook.

\section{The model}

\label{section:model} Our starting point is the Heisenberg Hamiltonian with
anisotropic terms that for a magnetic monolayer has the form
\cite{lit:rev_morpurgo}
\begin{equation}
H^{sl}=-\sum_{\langle i,j\rangle,\alpha}(J_{ij}\hspace{3pt}\vec{S}_{i}%
\cdot\vec{S}_{j}+\Lambda_{\alpha}S_{i}^{\left(  \alpha\right)  }S_{j}^{\left(
\alpha\right)  })-\sum_{i}A\left(  S_{i}^{\left(  z\right)  }\right)
^{2}\hspace{7pt}.\label{eq:spinmodel}%
\end{equation}
Here $J_{ij}$ is the exchange interaction between spins that favors
ferromagnetic $(J_{ij}>0)$ or antiferromagnetic $(J_{ij}<0)$ order of the
classical ground state, respectively. Because the exchange interaction is
short-ranged, that between nearest neighbors $\langle ij\rangle$ dominates,
while more distant ones can be disregarded. $A$ is the single-ion anisotropy
perpendicular to the plane, and $\Lambda_{\alpha}$ parameterizes an anisotropy
in the exchange interaction in a direction $\alpha$. These parameters depend
on the material and can be tuned externally such as by an applied
magnetic field or a gate voltage. In this paper we disregard the single-ion
anisotropy ($A=0$) but retain the anisotropic exchange assuming out-of-plane
anisotropy, $\Lambda_{z}=\Lambda$, $\Lambda_{x}=\Lambda_{y}=0$, noting that to
leading order $A$ and $\Lambda$ are equivalent. We disregard any spin-orbit interactions at this stage.

\section{Review of FM mono- and FM bilayers}

\label{section:ferro}

We first review the Holstein-Primakoff transformation, the method of choice to
treat the low frequency spectrum of spin Hamiltonians, as applied to FM
monolayers \cite{Owerre,fernandez-rossier} with isotropic exchange interaction
(\ref{generalmethod}). Afterwards, we review different types of anisotropy in
the FM coupling of the monolayer\cite{fernandez-rossier}
(\ref{mono:anisotropies}). Finally we consider a FM bilayer for isotropic
exchange coupling as well as out-of-plane anisotropy and review the dispersion
(\ref{FMbilayer}) \cite{Owerre}. This section serves essentially for fixing the geometry and the notation.

\subsection{General method}

\label{generalmethod}
\begin{figure}[th]
\centering
\includegraphics[width=0.4\textwidth]{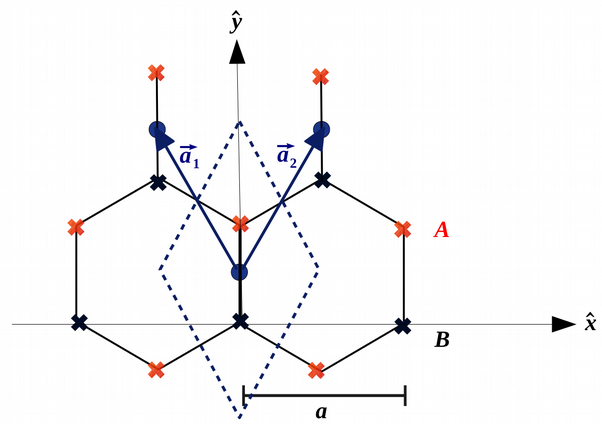}\hfill\caption[b]{Direct
triangular Bravais lattice with a two-atomic basis $A,B$ (crosses).Basis
vectors $\vec{a_{1}}$, $\vec{a_{2}}$ span the primitive unit cell as indicated
by dashed lines. Blue circles indicate lattice point and $a$ is the lattice
constant.}%
\label{fig1}%
\end{figure}

\begin{figure}[th]
\centering
\includegraphics[width=0.5\textwidth]{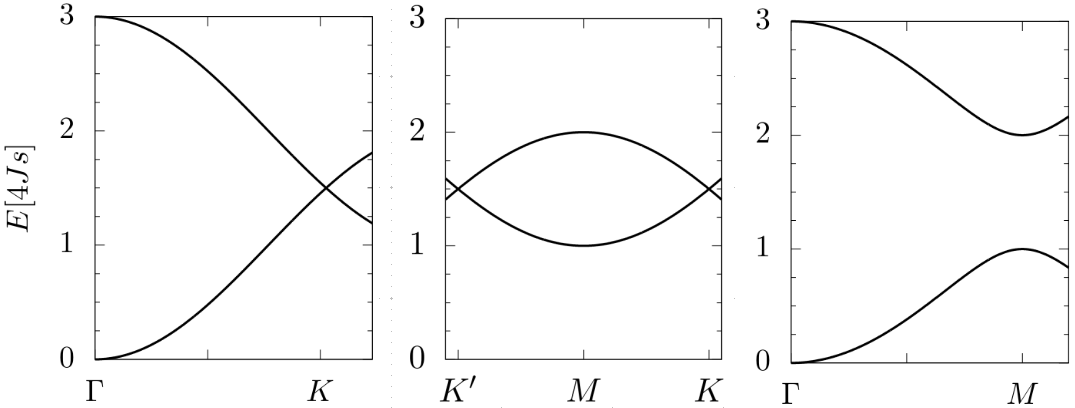}\hfill\caption{Energy
dispersion of an FM monolayer with isotropic exchange coupling along high
symmetry directions in the first BZ. $K$, $K^{\prime}$ are the inequivalent
Dirac points. }%
\label{fig2}%
\end{figure}

The Holstein-Primakoff (HP) transformation of the Hamiltonian (\ref{eq:spinmodel})
replaces the local spin operators $S_{j}$ in favor of Boson operators $a_{j}$
\cite{primakoff,stancil}:
\begin{align}
S_{j}^{+} &  =\sqrt{2s}\left(  1-\frac{a_{j}^{\dagger}a_{j}}{2s}\right)
^{1/2}a_{j}\ ,\nonumber\label{eq:HP}\\
S_{j}^{-} &  =\sqrt{2s}a_{j}^{\dagger}\left(  1-\frac{a_{j}^{\dagger}a_{j}%
}{2s}\right)  ^{1/2},\nonumber\\
S_{j}^{\left(  z\right)  } &  =s-a_{j}^{\dagger}a_{j}.
\end{align}
At low temperatures or weak excitation we may disregard all but the zeroth
order in $a/\sqrt{2s}$ in the series expansion of the square root. A single
boson excitation $\left\langle a^{+}a\right\rangle =1$ changes the spin
projection $\Delta S^{z}=\hbar$ parallel to the quantization axis $z$ and
perpendicular to the plane. After subtracting the constant ground state
energy, the Hamiltonian with FM exchange interaction and zero anisotropy
($\Lambda=A=0$) reads
\begin{equation}
H=-2Js\sum_{\langle i,j\rangle}a_{j}^{\dagger}a_{i}+2JsZ_{\mathrm{n.n}}%
\sum_{i}a_{i}^{\dagger}a_{i}.
\end{equation}
$Z_{\mathrm{n.n}}=3$ is the number of nearest neighbors of magnetic cations on
a hexagonal lattice. The lattice can be spanned by a triangular Bravais
lattice with a two-atomic basis (see Fig. ~\ref{fig1}). Transformation to
momentum space leads to non-interacting magnons
\begin{equation}
H=\sum_{k,r=\pm}\hbar\omega_{r,k}a_{r,k}^{\dagger}a_{r,k}%
\ ,\label{monolayer_magnon_Ham}%
\end{equation}
with energies\cite{fernandez-rossier}
\begin{equation}
E_{\pm}(k)=\hbar\omega_{\pm,k}=2Js\left(  3\pm|c_{k}|\right)  .
\end{equation}
Here $c_{k}=1+e^{-i\vec{k}\vec{a}_{1}}+e^{-i\vec{k}\vec{a}_{2}}$ is the
structure factor of the lattice with unit cell vectors $\vec{a}_{1},\vec
{a}_{2},$ as depicted in figure \ref{fig1}. This dispersion is isomorphic with
the $\pi$-electrons in monolayer graphene, as shown in Fig. \ref{fig2} for the
first BZ. It has a minimum and maximum at the $\Gamma$-point ($k=0$) and two
non-equivalent Dirac cones at the $K$ and $K^{\prime}$ corners at energy
$6Js$. with conical dispersion. 

\subsection{Anisotropies}

\label{mono:anisotropies} In CrI$_{3}$ \cite{lit:nanolett_cri3} the magnetic
anisotropy has an easy axis along $\hat{z}$, i.e. perpendicular to the plane
of the material. The Hamiltonian (\ref{eq:spinmodel}) becomes
\begin{equation}
\hat{H}\hspace{3pt}=-J\sum_{\langle i,j\rangle}\left(  S_{i}^{x}S_{j}%
^{x}+S_{i}^{y}S_{j}^{y}\right)  \hspace{3pt}-(J + \Lambda)\sum_{\langle
i,j\rangle}\hspace{3pt}S_{i}^{z}S_{j}^{z}\hspace{7pt}%
\end{equation}
with $\Lambda, J>0$. The dispersion \cite{fernandez-rossier}
\begin{equation}
E_{\pm}(k)=(6(J + \Lambda)s\pm2Js|c_{k}|)\ ,
\end{equation}
is shifted by $6\Lambda s$ compared to the isotropic case. This shift reflects
the suppression of the Goldstone mode of rotationally symmetric systems by
opening a \emph{spin wave gap} at the $\Gamma$-point. In the expansion of the
HP-transformation, we restricted to leading order, thereby neglecting
magnon-magnon interactions that become relevant at finite temperature. A
mean-field treatment of higher order bosonic operators renormalizes the
exchange coupling constants, and thereby also the spin wave
gap\cite{fernandez-rossier}.

\begin{figure}[th]
\centering
\includegraphics[width=0.5\textwidth]{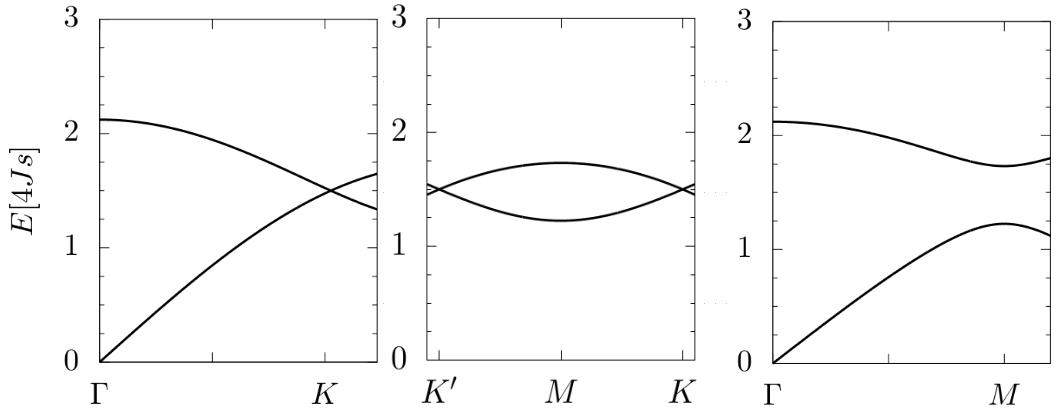}\hfill\caption{Energy dispersion for an FM monolayer with easy-plane anisotropy
$\Lambda=-J$. For further explanation see the text.}%
\label{fig3}
\end{figure}

We model an easy-plane anisotropic FM with $J>0$ and $\Lambda<0$ in the
Hamiltonian (\ref{eq:spinmodel}). We eliminate the non-bilinear terms of the
bosonic operators $a_{k}$
by a Bogoliubov transformation \cite{lit:kowalska}, which leads to quadratic
forms of Bose operators assigned to at most to two sublattices with spectrum:
\begin{align}
E_{\pm}  &  =Js\sqrt{R\pm S}%
\ ,\nonumber\label{monolayer_spectrum_with_anisotropy}\\
R  &  =36+4(1+\frac{\Lambda}{J})|c_{k}|^{2},\\
S  &  =24(1+\frac{\Lambda}{2J})|c_{k}|.\nonumber
\end{align}
$\Lambda=-J$ recovers the XY-model with dispersion \cite{lit:owerre}%
,\cite{fernandez-rossier}%
\begin{equation}
E_{\pm}=6Js\sqrt{1\pm\frac{|c_{k}|}{3}}\hspace{7pt}%
\end{equation}
plotted in Fig. \ref{fig3}. The general monolayer Hamiltonian
(\ref{monolayer_spectrum_with_anisotropy}) have recently studied in Ref.
\cite{Aguilera}. Note that $E_{\pm}$ is proportional to the square root of the
energy in the isotropic case. The easy-plane anisotropy was observed in a
monolayer of $\mathrm{CrCl}_{3}$ \cite{Soriano2020Sep,pinto,cai,wang}, which
should therefore be a good system to study phase transitions in 2D.
 
\subsection{FM bilayer}

\label{FMbilayer} 
\begin{figure}[th]
\centering
\includegraphics[width=0.45\textwidth]{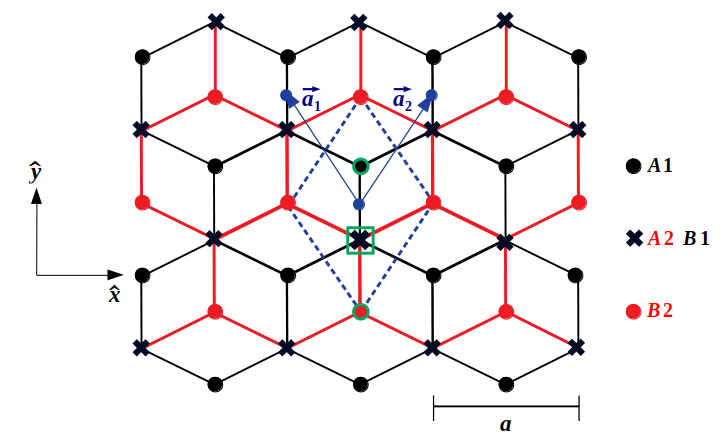}\hfill\caption{A bilayer
with $AB$-stacking as for example in bulk BiI$_{3}$ crystals. The primitive
unit cell (dashed blue lines) contains four atoms, $A1$ (green-rimmed black
dot) of bottom layer (label $1$), $B2$ (red-green) of top layer $2$ and the
stacked pair of atoms $B1$-$A2$ (black-green cross) with $A2$ on top of $B1$.
The basis vectors $\vec{a}_{1}$, $\vec{a}_{2}$ of the bilayer-lattice are the
same as for the monolayer and are shown as blue arrows.}%
\label{fig4}%
\end{figure}

For a bilayer with FM intra- and interlayer coupling
$J_{\parallel},J_{\perp}>0$ and without anisotropies we arrive at the
Hamiltonian
\begin{equation}
\hat{H}=-2J_{\Vert}\sum_{\langle i,j\rangle}\vec{S}_{i}\cdot\vec{S}%
_{j}-2J_{\bot}\sum_{\langle i,j\rangle}\vec{S}_{i}\cdot\vec{S}_{j}%
,\label{eq:bil_iso_fm}%
\end{equation}
where the first and second terms describe intra- and interlayer coupling,
respectively. We adopt the ratio of $J_{\perp}=0.26J_{\parallel}$ as predicted
for CrI$_{3}$ by first-principles calculations \cite{lit:nanolett_cri3}. We
consider here $AB$ type stacking of 2D hexagonal lattices with a lateral shift
by $[2/3,1/3]$ unit vectors (see Fig. \ref{fig4}) \cite{lit:nanolett_cri3},
which corresponds to the FM low-temperature crystallographic phase of bulk
$\mathrm{CrI}_{3}$ \cite{mcguire,soriano,jiang}. We chose a unit cell for a bilayer with four atoms, $A$-atoms $A1$
in the bottom-layer (1) and $A2$ in the top-layer (2) as well as $B$-atoms
$B1$ and $B2$ (see Fig. \ref{fig4}). Each $A$-($B$)-atom has three nearest
neighbors in the same layer belonging to the $B$-($A$)-sublattice. The atoms
$A2$ on top of $B1$ form another pair of nearest-neighbours per unit cell. The
magnon band structure consists now of four rather than two energy bands
\cite{Owerre}
\begin{align}
E_{\pm}^{[1]} &  =12J_{\parallel}s\pm4J_{\parallel}s|c_{k}|\\
E_{\pm}^{[2]} &  =12J_{\parallel}s+4J_{\perp}s\pm4s\sqrt{J_{\perp}%
^{2}+J_{\parallel}^{2}|c_{k}|^{2}}\hspace{7pt},
\end{align}
which reflects the more complex unit cell. The lowest band $E_{-}^{[1]}$ is
gapless at the origin because in the absence of any anisotropy the system is
invariant with respect to a global spin rotation. At the Dirac points $K$,
$K^{\prime}$, the structure factor vanishes and $E_{+}^{[2]}=(12sJ_{\parallel
}+8sJ_{\perp})$, $E_{1}=12sJ_{\parallel}$, where $E_{1}$ is threefold
degenerate. This spectrum differs from that of the $\pi$-electrons in bilayer
graphene, which are two-fold degenerate at the $K$ and $K^{\prime}$ points
with parabolic dispersion \cite{McCann}. The wave functions at the Dirac
points read%
\begin{align}
\Psi_{K^{(\prime)}}^{\left[  1\right]  } &  =\frac{1}{\sqrt{N}}\sum
_{j}e^{iK^{(\prime)}\vec{R}_{j}}a_{j,A1}^{\dagger}|0\rangle
,\label{eq:FMeigenstates2}\\
\Psi_{K^{(\prime)}}^{\left[  2\right]  } &  =\frac{1}{\sqrt{N}}\sum
_{j}e^{iK^{(\prime)}\vec{R}_{j}}a_{j,B2}^{\dagger}|0\rangle,\\
\Psi_{K^{(\prime)}}^{\left[  3\right]  } &  =\frac{1}{\sqrt{2N}}\sum
_{j}e^{iK^{(\prime)}\vec{R}_{j}}\left(  a_{j,B1}^{\dagger}+a_{j,A2}^{\dagger}\right)
|0\rangle,\\
\Psi_{K^{(\prime)}}^{\left[  4\right]  } &  =\frac{1}{\sqrt{2N}}\sum
_{j}e^{iK^{(\prime)}\vec{R}_{j}}\left(  a_{j,B1}^{\dagger}-a_{j,A2}^{\dagger}\right)
|0\rangle,\label{eq:FMeigenstates3}%
\end{align}
where $|j\alpha\rangle$ denotes the position of the site on sublattice
$\alpha$. The eigenstate $\Psi_{K^{(\prime)}}^{\left[  1\right]  }%
(\Psi_{K^{(\prime)}}^{\left[  2\right]  })$ is localized to sublattice
$A1\left(  B2\right)  $ in layer $1$($2).$ The sublattices $A2$ and $B1$ are
coupled by $J_{\bot}$, which generates an in phase or acoustic mode
$\Psi^{\left[  3\right]  }$ with lower energy $E_{1}$ or out-of-phase $\pi
$-shifted optical mode $\Psi^{\left[  4\right]  }$ at higher energy
$E_{+}^{\left[  2\right]  }$.

$\Psi^{\left[  1\right]  },\Psi^{\left[  2\right]  },\Psi^{\left[  3\right]
}$ correspond to excitations in which the spins on the same sublattice and
layer, separated by $a$ along $(1,0)$ or $(-\frac{1}{2},\frac{\sqrt{3}}{2}),$
precess with a relative phase shift $\frac{2\pi}{3}$. The spin precession
therefore reflects the structure of the hexagonal lattice bonds at the
Dirac-points. In the Appendix we demonstrate that these modes also solve the
Landau-Lifshitz equation for coupled classical spins.

A perpendicular anisotropy can be modeled by the coupling constants
$J_{\parallel}^{\left(  zz\right)  }$, $J_{\perp}^{\left(  zz\right)  }$ and
blue shifts the frequencies,\begin{widetext}
\begin{align}
E_{\pm}^{[1]} \hspace{3pt} &= \hspace{3pt} 12sJ_{\|}^{zz} \hspace{3pt} + \hspace{3pt} 2s(J_{\bot}^{zz}-J_{\bot}) \hspace{3pt} \pm \hspace{3pt} 2s\sqrt{(J_{\bot}^{zz}-J_{\bot})^2 \hspace{3pt} + \hspace{3pt} 4 J_{\|}^{2} \vert c_k \vert^2}& \\
E_{\pm}^{[2]} \hspace{3pt} &= \hspace{3pt} 12sJ_{\|}^{zz} \hspace{3pt} + \hspace{3pt} 2s(J_{\bot}^{zz}+J_{\bot}) \hspace{3pt} \pm \hspace{3pt} 2s\sqrt{(J_{\bot}^{zz}+J_{\bot})^2 \hspace{3pt} + \hspace{3pt} 4 J_{\|}^{2} \vert c_k \vert^2}&
\end{align}
\end{widetext}Moreover, the triple degeneracy at the Dirac points is reduced
to a double one.

\section{Bilayer with FM intra- and AFM interlayer exchange coupling}

\label{section:antiferro} We first derive new results for the dispersion for a
bilayer with FM intralayer and AFM interlayer isotropic exchange interactions
(\ref{AFMisotropic}). Subsequently, we include a perpendicular-plane
anisotropy and focus on the analysis of the fundamental gap at $\Gamma$
(\ref{AFManisotropic}). The topology in terms of the Berry curvature is subject of Section \ref{AFMtopology}.

\subsection{Isotropic exchange interaction}

\label{AFMisotropic} Several papers discuss the impact of stacking \cite{Soriano2020Sep,soriano, lit:nanolett_cri3, jiang, jang} on interlayer magnetic
coupling of a CrI$_3$ bilayer. Depending on the type of involved
interlayer orbital hybridizations, the corresponding coupling
of the modeling spin Hamiltonian is FM or AFM type. For AB
stacking, it has been shown by density functional theory calculations [19] that both NN and NNN interactions determine
the order of the bilayer ground state: There are one NN neighbor and 16 NNN within a unit cell, the NN contributing with
AFM coupling whereas the NNN contributing with FM coupling, so that in total, interlayer magnetism in AB stacking is
strongly FM.   As magnetic interlayer order can be tuned by application of an electrostatic gate \cite{huang} or a magnetic field \cite{lit:ref39}, however, we find it instructive to discuss both types  of interlayer coupling (FM and AFM) for the same type of stacking. Here we choose AB-stacking for simplicity 
and an AFM interlayer magnetism of the bilayer, as is induced by the NN couplings of the AB-stacking.

We first calculate the energy dispersion of a bilayer
with isotropic exchange coupling for different spin directions and
intra/interlayer coupling strengths $J_{\parallel}/J_{\perp}$. The Hamiltonian
reads with $J_{\parallel},J_{\perp}>0$%

\begin{equation}
\hat{H}=-2J_{\parallel}\sum_{\langle i,j\rangle\in\{\mathrm{intra}\}}\vec
{S}_{i}\cdot\vec{S}_{j}+2J_{\perp}\sum_{\langle i,j\rangle\in\{\mathrm{inter}%
\}}\vec{S}_{i}\cdot\vec{S}_{j}.\label{eq:AFM_iso1}%
\end{equation}
Again, the first sum includes the three in-plane nearest neighbors of a local
moment on site $i$, while the second sum runs over closely spaced dimers
$A2$,$B1$ between the layers. 
When $S_{z}=s$ for the spins in the top layer ($2$), $S_{z}=-s$ in the bottom layer
($1$) minimizes the classical ground state energy $E_{0}$. The magnons
$a_{i}^{+},a_{i}$ are the excitations. We apply the HP-
transformation and expand Eq. (\ref{eq:AFM_iso1}) to leading order in the
magnon operators, thereby disregarding magnon-magnon interactions, which is
valid at low temperatures. In a mean-field approximation, higher terms only
renormalize the exchange constants \cite{arakawa}, as
confirmed by experiments work on bilayer CrI$_{3}$ \cite{lit:ref39},
at a temperature $T=0.033\,J$. Therefore%
\begin{align}
S_{i,\alpha2}^{+(-)} &  =\sqrt{2s}a_{i,\alpha2}^{(+)},\;S_{i,\alpha2}%
^{z}=s-a_{i,\alpha_{2}}^{+}a_{i,\alpha2},\\
S_{i,\alpha1}^{-(+)} &  =\sqrt{2s}a_{i,\alpha1}^{(+)},\;S_{i,\alpha1}%
^{z}=-s+a_{i,\alpha1}^{+}a_{i,\alpha1}.
\end{align}
The subscripts refer to atom $\alpha\in\{A,B\}$ of lattice cell $i$ in layer
$\nu\in\{1,2\}$. The magnon Hamiltonian then reads \begin{widetext}
\begin{eqnarray} \label{eq:AFM_iso_bos_real}
\hat{H} - E_0  = & -2 s J_{\parallel}  \sum_{\langle (i,\alpha \nu), (j,\alpha^{\prime} \nu)\rangle, \alpha \neq \alpha^{\prime}}  (a_{j,\alpha^{\prime} \nu}^{+} a_{i,\alpha \nu}  +  h.c.)  +  6sJ_{\parallel}  \sum_{i,\alpha \nu} a_{i,\alpha \nu}^{+} a_{i,\alpha \nu}& \nonumber \\
&+ 2sJ_{\perp} \sum_i (a_{i,A2}^{+}a_{i,B1}^{+}  + h.c.)  +  2sJ_{\perp}  \sum_i   (a_{i,A2}^{+} a_{i,A2}   +
a_{i,B1}^{+}a_{i,B1}).
\end{eqnarray}
\end{widetext}As common for antiferromagnetic order, the classical ground
state is not an eigenstate of the Hamiltonian since
\begin{equation}
a_{i,A2}^{+}a_{i,B1}^{+}|\downarrow\rangle_{1}|\uparrow\rangle_{2}%
\propto|\uparrow\rangle_{1}|\downarrow\rangle_{2}\hspace{3pt}\neq0.
\end{equation}
We can accommodate this issue by writing the Hamiltonian in reciprocal space
as \cite{lit:colpa}
\begin{equation}
\hat{H}-E_{0}=E_{c}+\sum_{\vec{k}}(\vec{a}_{\vec{k}}^{+},\vec{a}_{-\vec{k}%
})\mathfrak{D}(\vec{a}_{\vec{k}},\vec{a}_{-\vec{k}}^{+})^{T}\label{eq:HwithD}%
\end{equation}
where $\vec{a}_{\vec{k}}=(a_{\vec{k},A1},a_{\vec{k},B1},a_{\vec{k},A2}%
,a_{\vec{k},B2})$, $E_{c}$ a constant to be discussed later, and
$\mathfrak{D}$ is the $8\times8$-matrix
\begin{equation}
\mathfrak{D}=%
\begin{pmatrix}
A & B\\
B & A
\end{pmatrix}
\end{equation}
in which
\begin{gather}
A=%
\begin{pmatrix}%
\begin{matrix}
3\hspace{1pt}J_{\parallel}s & -J_{\parallel}sc_k^{\ast}\\
-J_{\parallel}s\hspace{3pt}c_k & 3J_{\parallel}s+J_{\perp}s
\end{matrix}
& \vline & 0\\\hline
0 & \vline &
\begin{matrix}
3\hspace{1pt}J_{\parallel}s+J_{\perp}s & -J_{\parallel}sc_k^{\ast}\\
-J_{\parallel}sc_k & 3J_{\parallel}s
\end{matrix}
\end{pmatrix}
,\label{matrix:iso_AFM}\\
B=%
\begin{pmatrix}
&  & 0 &  & \\\hline
0 & \vline &
\begin{matrix}
0 & J_{\perp}s\\
J_{\perp}s & 0
\end{matrix}
& \vline & 0\\\hline
&  & 0 &  &
\end{pmatrix}
,
\end{gather}
and $c_k$ is again the structure factor of the hexagonal lattice. Kowalska's
framework \cite{lit:kowalska} is not applicable for four sublattices. Instead, we
diagonalize the Hamiltonian
by a para-unitary transformation $\mathfrak{T}$ of operators $(\vec{a}%
_{\vec{k}},\vec{a}_{-\vec{k}}^{+})^{T}$ to the bosonic operators $\vec{\gamma
}_{\vec{k}}$ \cite{lit:colpa}:
\begin{equation}
(\vec{\gamma}_{\vec{k}},\vec{\gamma}_{-\vec{k}}^{+})^{T}=\mathfrak{T}(\vec
{a}_{\vec{k}},\vec{a}_{-\vec{k}}^{+})^{T}%
\end{equation}
such that
\begin{align}
\hat{H}-E_{0}-E_{c} &  =\sum_{\vec{k}}(\vec{a}_{\vec{k}}^{+},\vec{a}_{-\vec
{k}})\mathfrak{T}^{\dagger}(\mathfrak{T}^{\dagger})^{-1}\mathfrak{D}%
\mathfrak{T}^{-1}\mathfrak{T}(\vec{a}_{\vec{k}}^{+},\vec{a}_{-\vec{k}%
})^{\dagger}\nonumber\\
&  =\hbar\sum_{\vec{k}}(\vec{\gamma}_{\vec{k}}^{+},\vec{\gamma}_{-\vec{k}%
})\mathrm{diag}(\omega_{1},..,\omega_{4},\omega_{1},..,\omega_{4})(\vec
{\gamma}_{\vec{k}}^{+},\vec{\gamma}_{-\vec{k}})^{\dagger}\nonumber\\
&  =2\sum_{\vec{k},r=1}^{4}\hbar\omega_{r}\left(  \gamma_{r,\vec{k}}^{+}%
\gamma_{r,\vec{k}}+\frac{1}{2}\right)  ,\label{eq:colpa_transform_schemme}%
\end{align}
provided that $\mathfrak{D}$ is positive-definite. $E_{c}$ is a further
constant that will be specified below. $\mathfrak{T}$ is para-unitary in the
sense that
\begin{equation}
\mathfrak{T}\eta\mathfrak{T}^{\dagger}=\eta,
\end{equation}
with $\eta=\mathrm{diag}(I_{4},-I_{4}),$ where $I_{n}$ is the unit matrix with
dimension $n$, which ensures that the $\gamma_{r,\vec{k}}^{(+)}$ obey bosonic
commutation relations. $(\lambda_{1},...,\lambda_{8}):=(\omega_{1}%
,..,\omega_{4},-\omega_{1},..,-\omega_{4})$ are the para-values of
$\mathfrak{D}$
\begin{equation}
(\mathfrak{D}-\lambda_{i}\eta)\vec{v}_{i}=0\label{eq:paraw_problem}%
\end{equation}
with para-vectors $\vec{v}_{i}$. Eq. (\ref{eq:paraw_problem}) can be written
as an eigenvalue problem by multiplying by $\eta$ from the left
\begin{equation}
(\eta\mathfrak{D}-\lambda_{i}I)\vec{v}_{i}=0.\label{eq:eigenw_problem}%
\end{equation}
Diagonalizing the non-Hermitian matrix $\eta\mathfrak{D}$ leads to a set of
four positive and four negative eigenvalues $\pm\lambda_{i}$ corresponding to
the two twofold degenerate energy bands \begin{widetext}
\begin{equation}\label{eq:dispAFMiso}
E_{\pm} = s  \sqrt{3 J_{\parallel} J_{\perp} + 9 J_{\parallel}^2 \left( 1 + \frac{\vert c_k\vert^2}{9} \right)  \pm  \sqrt{3} J_{\parallel} \sqrt{3 J_{\perp}^2 + (12 J_{\parallel}^2 + 4 J_{\parallel}J_{\perp})\vert c_k \vert^2}}
\hspace{7pt}.
\end{equation}
\end{widetext}

The difference in energy bands for bilayers with AFM and FM order can be traced to the matrix $\eta$. In physical terms, two AFM-coupled sublattices ($A2$-$B1$) generate two mode families that are exchanged by a $\pi$-rotation of the bilayer and hence are degenerate. The additional symmetry is also responsible for the degenerate ground state of the AFM bilayer. Breaking the interlayer symmetry by perpendicular electric and magnetic fields removes the degeneracy \cite{Owerre}. 

\begin{figure}[th]
\centering
\includegraphics[width=0.5\textwidth]{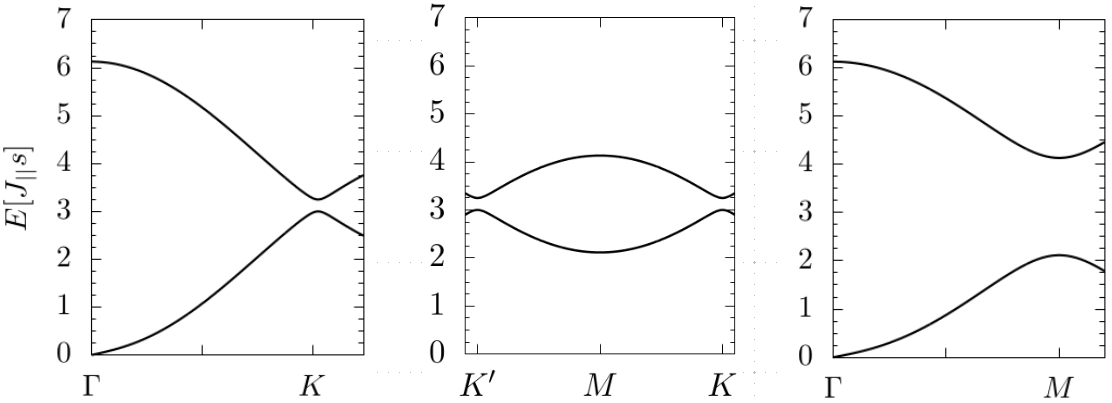}\hfill\caption{Dispersion
of a bilayer with AFM inter-layer and FM intra-layer coupling, isotropic
exchange coupling constants and a ratio of inter- vs. intralayer coupling
$J_{\perp}=0.26\hspace{2pt}J_{\parallel}$. We observe a gap of order
$J_{\perp}s$ at the Dirac points with quadratic instead of the linear
dispersion of the FM monolayer in Fig. \ref{fig2}.}%
\label{fig5}%
\end{figure}

The dispersion (\ref{eq:dispAFMiso}) is plotted in Fig. \ref{fig5}. We find a
difference $\triangle E$ between the the zero-point energy of the magnon
system and the classical ground state energy $E_{0}=-12NJ_{\parallel}%
s^{2}-2J_{\perp}Ns^{2}$
\begin{equation}
\triangle E=E_{c}+\sum_{\vec{k}}\sum_{r=1}^{4}\hbar\omega_{r}%
=-Ns(12J_{\parallel}+2J_{\perp})+\sum_{\vec{k}}\sum_{r=1}^{4}\hbar\omega_{r},
\end{equation}
see also Eq. (\ref{eq:colpa_transform_schemme}). The first term on the right
hand side $E_{c}=E_{0}/s$, arises from quantum fluctuations of the
$z$-component, while the second term reflects transverse fluctuations cause.
In the following we disregard these zero-point fluctuations, but recommend
their study in a future project.

Around the Dirac-points $K,K^{\prime}$, the dispersion can be expanded up to
second order in $k$ as
\begin{align}
E_{+}(k) &  =\sqrt{3}J_{\parallel}s\hspace{2pt}\sqrt{3+2\frac{J_{\perp}%
}{J_{\parallel}}}+\frac{3}{8}a^{2}J_{\parallel}s\frac{3+6\frac{J_{\parallel}%
}{J_{\perp}}}{\sqrt{9+6\frac{J_{\perp}}{J_{\parallel}}}}k^{2},\nonumber\\
E_{-}(k) &  =3J_{\parallel}s-\frac{1}{8}a^{2}J_{\parallel}s\left(
1+6\frac{J_{\parallel}}{J_{\perp}}\right)  k^{2}.
\end{align}
The AFM coupling $J_{\perp}$ therefore opens a gap of the order $sJ_{\perp}$
at $K$, $K^{\prime}$, leading to a quadratic rather than the linear dispersion
found for the FM monolayer, but different effective masses. This gap implies a
possible non-trivial topology. However, the Chern numbers are found to be zero for each branch, which we indicate in Section \ref{sec:topology}.
 
\subsection{Anisotropy}

\label{AFManisotropic} 
\begin{figure}[th]
\centering
\includegraphics[width=0.5\textwidth]{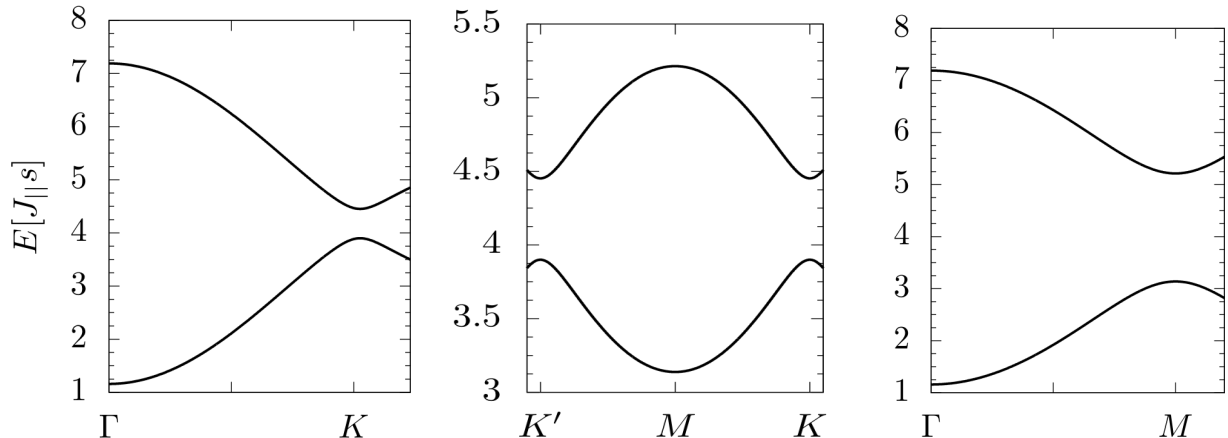}\hfill\caption{Magnon
dispersion of a bilayer with AFM inter-layer and FM intra-layer coupling with
anisotropic exchange coupling $J^{zz}\neq J^{xx}=J^{yy}=J$. Here the
inter-layer couplings $J_{\perp}=0.26\hspace{2pt}J_{\parallel}$, $J_{\perp
}^{zz}=0.56\hspace{2pt}J_{\parallel}$ and intra-layer coupling $J_{\parallel
}^{zz}=1.3\hspace{2pt}J_{\parallel}$.}%
\label{fig6}%
\end{figure}

Next, we introduce an out-of-plane anisotropy with
$J_{\parallel}^{zz}>J_{\parallel}$,$J_{\perp}^{zz}>J_{\perp}$. The matrix $A$
then reads
\[
A=%
\begin{pmatrix}%
\begin{matrix}
3J_{\parallel}^{zz}s & -J_{\parallel}sc_k^{\ast}\\
-J_{\parallel}sc_k & 3\hspace{1pt}J_{\parallel}^{zz}s+J_{\perp}^{zz}s
\end{matrix}
& \vline & 0\\\hline
0 & \vline &
\begin{matrix}
3J_{\parallel}^{zz}s+J_{\perp}^{zz}s & -J_{\parallel}sc_k^{\ast}\\
-J_{\parallel}sc_k & 3J_{\parallel}^{zz}s
\end{matrix}
\end{pmatrix}
,
\]
while $B$ is not affected. We can still derive an analytic expression for the
energy dispersion \begin{widetext}
\begin{equation} \label{eq:afm_aniso}
E_{\pm} = \frac{s}{\sqrt{2}} \sqrt{18 J_{\parallel}^{zz^2} + 6 J_{\parallel}^{zz}J_{\perp}^{zz} + J_{\perp}^{zz^2} - J_{\perp}^2 + 2 J_{\parallel}^2 \vert c_k \vert^2
\pm  \sqrt{(6 J_{\parallel}^{zz}J_{\perp}^{zz} + J_{\perp}^{zz^2} - J_{\perp}^2 )^2  +  [(12 J_{\parallel}^{zz} + 2 J_{\perp}^{zz})^2 - 4 J_{\perp}^2] J_{\parallel}^2 \vert c_k \vert^2}} .
\end{equation}
\end{widetext}and plot them in Figure \ref{fig6} for coupling constants
$J_{\parallel}^{zz}=1.3J_{\parallel},$ $J_{\perp}^{zz}=0.56\hspace
{2pt}J_{\parallel},$ $J_{\perp}=0.26\hspace{2pt}J_{\parallel}$. Here we adopt
again a ratio of $0.26$ between inter- and intra layer coupling.
We assume that FM and AFM ordered layers are both AB stacked and that the ratio between inter and intra-layer coupling ($0.26$ for FM CrI$_{3}$ \cite{lit:nanolett_cri3}) only changes sign. Actually, AFM ordered CrI$_{3}$ has both a different ($AB^{\prime}$) stacking and the interlayer exchange is smaller with an inter/intra layer ratio of $-0.018$. Other constants are known for monolayer CrI$_{3}$
\cite{kitaev,fernandez-rossier} and can be tuned, for example, by an
electrostatic gate.\cite{huang}. Here we chose them to enhance the visibility
of the effects in the figures. The anisotropy blue-shifts the lower band edge $\sim J_{\parallel
}s$ relative to the zero-point energy $E_{0}-Ns\hspace{2pt}(12J_{\parallel
}^{zz}+2J_{\perp}^{zz})+\sum_{\vec{k}=1}^{N}\sum_{r=1}^{4}\hbar\omega_{r}$ and
increases the gap at the Dirac points ($\sim J_{\perp}s$ for the isotropic
AFM-bilayer) to $\sim J_{\perp}^{zz}s$.
\begin{figure}[th]
\centering
\includegraphics[width=0.5\textwidth]{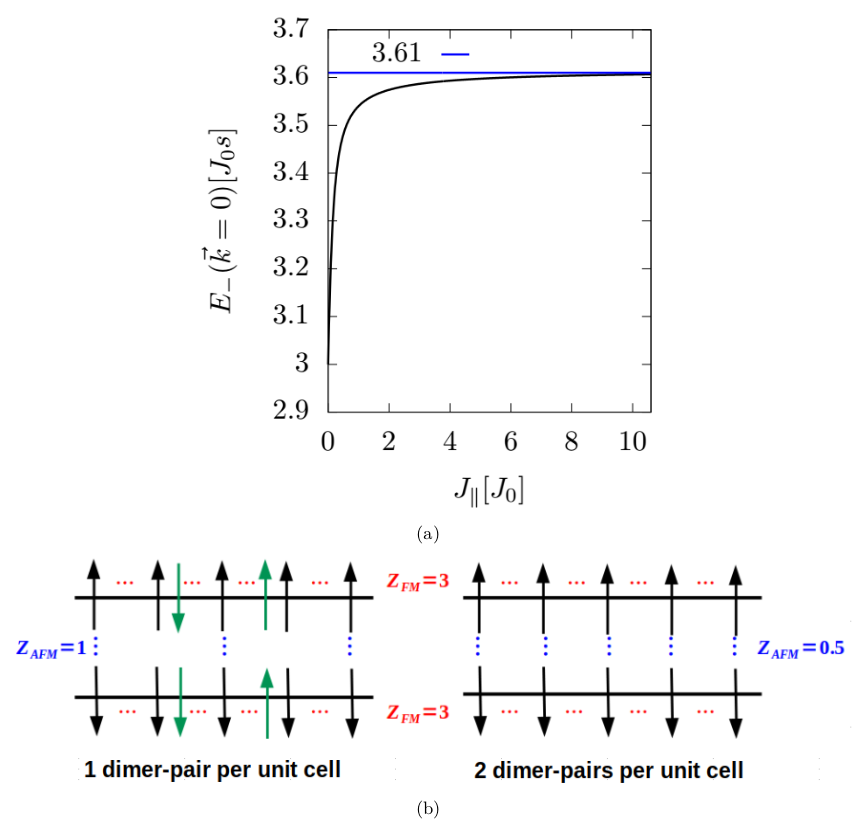}\hfill
\caption{(a) Magnon gaps for a realistic (black line) and a hypothetical
bilayer (blue line) as a function of FM coupling strength $J_{\parallel}$. The
anisotropy is constant with $J_{\parallel}^{zz}-J_{\parallel}=1.0\hspace
{2pt}J_{0}$, $J_{\perp}^{zz}-J_{\perp}=0.3\hspace{2pt}J_{0}$ and $J_{\perp
}=1.0\hspace{2pt}J_{0}$. (b)(left) Realistic bilayer schematic with
coordination numbers $Z_{AFM}=1$ and $Z_{FM}=3$. (b)(right) Hypothetical
bilayer schematic with $Z_{AFM}=0.5$ and $Z_{FM}=3$.}
\label{fig7}%
\end{figure}

We now analyze the fundamental gap $\hbar\omega_{-}(\vec{k}=0)$ (see Eq.
(\ref{eq:afm_aniso})) plotted in Figure \ref{fig7}(a) as a function of the FM
coupling strength $J_{\parallel}$ for $J_{\perp}=1.0\hspace{2pt}J_{0}%
$,$J_{\parallel}^{zz}-J_{\parallel}=1.0\hspace{2pt}J_{0}$ and $J_{\perp}%
^{zz}-J_{\perp}=0.3\hspace{2pt}J_{0}$. In a simple FM the gap
\begin{equation}
\Delta_{FM}\propto s(J_{\parallel}^{zz}-J_{\parallel})\hspace{7pt}%
\label{eq:deltafm}%
\end{equation}
depends on $J_{\parallel}$ only via anisotropy. The anisotropy gap in a pure AFM, on the
other hand,
\begin{equation}
\Delta_{AFM}\propto s\sqrt{(J_{\perp}^{zz}-J_{\perp})(J_{\perp}^{zz}-J_{\perp
}+2J_{\perp})}\label{eq:deltaafm}%
\end{equation}
depends not only on the anisotropy $J_{\perp}^{zz}-J_{\perp}$, but also on the
AFM coupling strength $J_{\perp}$ \cite{lit:nolting}. The increase of the intra-layer FM coupling
increases the gap $E_{-}(\vec{k}=0)$ according to Eq.(\ref{eq:afm_aniso}),
which by the reduced number of thermal magnons is equivalent to an enhanced
AFM coupling. 

We analyze this effect by computing the gap of a hypothetical structure in
which the contributions from Eq. (\ref{eq:deltafm}) of the FM and Eq.
(\ref{eq:deltaafm}) of the AFM coupling at $\vec{k}=0$ are clearly separated.
The stacking of two ferromagnetic monolayers in this \textquotedblleft bilayer
(II)\textquotedblright\ is slightly shifted such that there are two
AFM-coupled dimer pairs $A2-B1$ and $A1-B2$ with coordination number
$Z_{AFM}=0.5$ (see Figure \ref{fig7}b (right)) compared to the original
coordination number $Z_{AFM}=1$ for the single dimer-pair in bilayer (I) (see
Figure \ref{fig7}b (left)). The gap of this modified system \begin{widetext}
\begin{equation} \label{eq:2dimers}
E_-(k = 0) = s \sqrt{((J_{\parallel}^{zz} - J_{\parallel}) Z_{FM} \hspace{3pt} + \hspace{3pt} (J_{\perp}^{zz} - J_{\perp}) Z_{AFM}) \\\
((J_{\parallel}^{zz} - J_{\parallel}) Z_{FM} \hspace{3pt} + \hspace{3pt} (J_{\perp}^{zz} - J_{\perp}) Z_{AFM} \hspace{3pt} + \hspace{3pt} 2 J_{\perp} Z_{AFM})} \hspace{7pt}
\end{equation}
\end{widetext}does not depend explicitly on $J_{\parallel}$, but on
$J_{\parallel}^{zz}-J_{\parallel}$ , see Figure \ref{fig7}(a) (blue line) and
Eq.(\ref{eq:2dimers}). For $J_{\parallel}=0$, the gap $3J_{0}s=s(J_{\parallel
}^{zz}-J_{\parallel})Z_{FM}$ of bilayer (I) is governed by the anisotropy of
the FM intralayer exchange only, while the AFM coupling does not contribute to
the gap. The gaps converge to $\sim3.61\hspace{2pt}J_{0}s$ only when the FM
coupling in bilayer (I) $J_{\parallel}\gtrapprox5J_{\perp}$. This result
suggests that a strong FM intra-layer coupling in the realistic structure (I)
increases the AFM inter-layer coupling, while in the limit of weak FM
coupling, the AFM order of the classical GS is less stable than in bilayer
(II) (see green arrows in Figure \ref{fig7}(b)).

\begin{figure}[ht]
\centering
\includegraphics[width=0.5\textwidth]{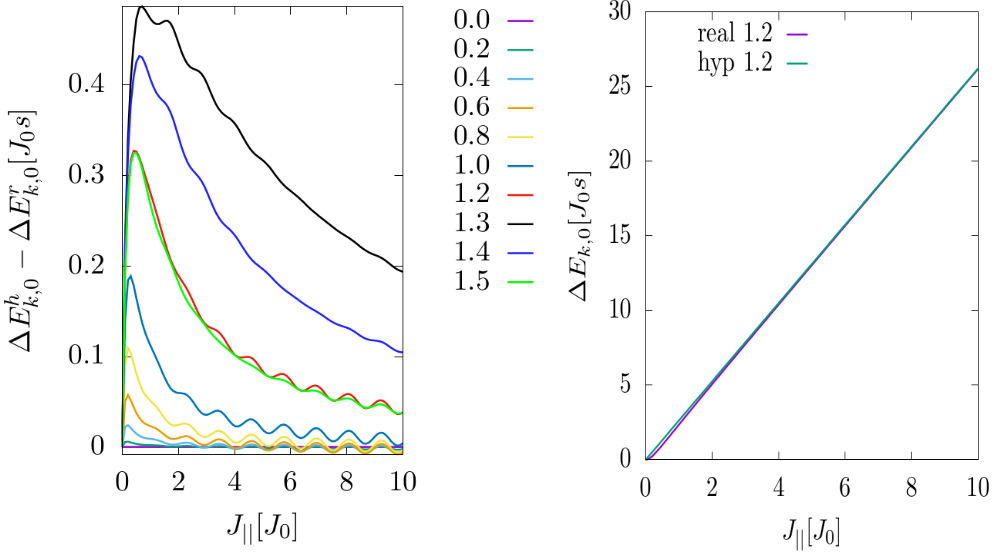}\hfill\caption{(Left)
Difference $\Delta E_{k,0}^{h}-\Delta E_{k,0}^{r}$ between the hypothetical
and a realistic bilayer structure as a function of FM coupling as a function
of $(k_{x},0)[\frac{\pi}{a}]$ along the $\Gamma-K$ direction in the first BZ.
(Right) Energy difference $\Delta E_{k,0}^{-}$ between a magnon with
wavevector $(1.2,0)[\frac{\pi}{a}]$ and zero wavevector in the lower band as a
function of FM coupling strength $J_{\parallel}$ for bilayers I (green) and II
(violet). }%
\label{fig:AFMstrength}%
\end{figure}

This statement is corroborated by the finite-wave vector magnon dispersion
$\Delta E_{k,0}=E_{-}(\vec{k})-E_{-}(0)$ as a function of the FM coupling. The zero-$k$-magnon is that of an interlayer AFM in its classical
GS. As $\Delta E_{k,0}$ measures the energy cost of exciting a finite-$k$-magnon, it thereby measures the AFM coupling strength. The right panel of figure \ref{fig:AFMstrength}
shows an $\Delta E_{k,0}$, which indeed increases with $J_{\parallel}$ for both
bilayers (I) and (II). The left panel of figure \ref{fig:AFMstrength} shows the difference
$\Delta E_{k,0}^{h}-\Delta E_{k,0}^{r}$ of a hypothetical and a real bilayer
for different points along the $\Gamma-K$ direction in the first BZ, which
decreases with increasing $J_{\parallel}$, confirming that the real bilayer
approaches the effective AFM coupling strength of the hypothetical bilayer for
large $J_{\parallel}.$ This shows that in the limit of strong intralayer coupling, magnetic order does no longer depend on the choice of stacking in our specific case.

\subsection{Topology}\label{sec:topology}

\label{AFMtopology} \begin{figure}[th]
\centering
\includegraphics[width=0.45\textwidth]{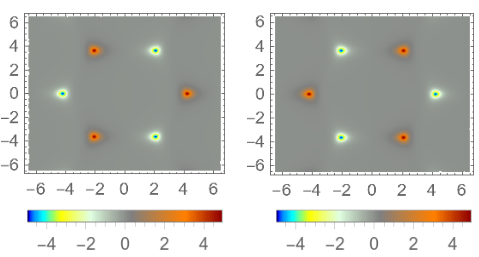}\hfill\caption{The Berry
curvature $\Omega_{xy;n}(\mathbf{k})$ for the bands $E_{+}$ (left) and $E_{-}$
(right) of a bilayer with AFM interlayer and FM intralayer coupling. The
exchange coupling constants are chosen as in Fig. \ref{fig6}}%
\label{fig8}%
\end{figure}

The topology of the magnon spectrum is reflected by the Berry curvature
$\Omega_{n\mathbf{k}}=\nabla_{\mathbf{k}}\times\langle u_{n\mathbf{k}}%
|i\nabla_{\mathbf{k}}|u_{n\mathbf{k}}\rangle$ of the $n=\pm$ bands
(\ref{eq:afm_aniso}), where $u_{n\mathbf{k}}$ is the periodic (Bloch) part of
the wave function \cite{Qiao}. For a Dirac-like spectrum, the Berry curvature
is large in the vicinity of the Dirac points, which dominate the topological
properties \cite{Zhai} as illustrated by Fig. \ref{fig8}. Their signs are
opposite at Dirac points $K$, $K^{\prime}$, which means that the Chern number
vanishes for each band. The topology for the bilayers in the anisotropic
exchange model without spin-orbit interaction is therefore trivial, without
protected edge states inside the gap. The thermal Hall conductivity, which is often used to probe topological properties of systems with a Dirac-like spectrum, is
proportional to the product of the Bose distribution function times the Berry
curvature $\Omega_{xy;n}(\mathbf{k})$ integrated over the first BZ
\cite{Hall} and vanishes as well.

This corresponds to the general fact that a non-vanishing thermal Hall
conductivity has so far been predicted for CrI$_{3}$ monolayer systems with
the anisotropy contributions to the spin Hamiltonian of the Kitaev model
\cite{Aguilera} or the DMI \cite{chen,lit:owerre}. More generally, Costa et al
\cite{costa} described magnons in monolayer CrI$_{3}$ by an itinerant fermion
model based on first-principles calculations, thereby circumventing model
assumptions for the anisotropy. They showed that the spin-orbit coupling of
iodine is essential for a non-trivial topology.

\section{Conclusions}

\label{section:conclusions} We report analytical expressions for the magnon
band structure of bilayers of two-dimensional ferromagnets with
(anti-) ferromagnetic interlayer exchange coupling and perpendicular anisotropy,
complementing previous numerical analysis \cite{Owerre}. An analytic
expression for the fundamental gap reveals AFM and FM contributions that can
be modeled by an effective coordination number. As the comparison of the
spectral properties between our real bilayer system and the hypothetical toy
model have shown, an increasing FM coupling in the real bilayer leads
effectively to a stronger AFM interlayer coupling. The spectral properties
refer to the analysis of the spectral gap as well as the energy cost
associated with adding an additional magnon to the system. Both results agree
with respect to the effect of stronger AFM coupling.

A natural extension of the present work would be to include
next-nearest-neighbour exchange interactions, which have been shown to have an impact on magnetic interlayer coupling \cite{lit:nanolett_cri3} for the $AB$-type stacking considered in this work.
We have shown that the Chern number vanishes in the exchange-anisotropy spin
model considered here, so that there is no magnon thermal Hall effect in the
absence of spin-orbit interaction or complex spin texture\textit{.}

\vskip0.25cm

\begin{acknowledgments}
L.O. acknowledges support by the DFG (RTG 1995). G.B. is supported by JSPS
KAKENHI Grant No. 19H006450.
\end{acknowledgments}

\section{Appendix: Classical consideration of FM-bilayer eigenmodes}

Here we show that the magnon modes at the Dirac points can be derived from a
purely classical torque cancellation argument.

Central to the Landau-Lifshitz equation is the torque $\vec{\tau}$ experienced
by a spin by a magnetic field $\vec{H}$:
\begin{equation}
\vec{\tau}=\frac{d\vec{S}_{i}}{dt}=\gamma\mu_{0}\vec{S}_{i}\times\vec{H},
\label{eq:torque}%
\end{equation}
where $\gamma= -g \mu_{B} < 0$ is the gyromagnetic ratio for the electron and
$\mu_{0}$ the permeability of free space. The coupling to neighboring spins
can be taken into account by an effective field $\vec{H}_{\mathrm{eff}}$
\cite{stancil}
\begin{equation}
\vec{H}_{\mathrm{eff}}=-\frac{2}{g\mu_{0}\mu_{B}}\sum_{j\in<i>}J_{ij}\vec
{S}_{j},
\end{equation}
where $\mu_{B}$ is the Bohr magneton and $g$ the Land\'e-factor. Then
\begin{equation}
\frac{d\vec{S}_{i}}{dt}=\gamma\mu_{0}\vec{S}_{i}\times\vec{H}_{\mathrm{eff}},
\end{equation}
When a spin belongs to a classical ground state that does not precess, the
torques cancel
\begin{align}
0  &  =J_{\parallel}\hspace{3pt}\vec{S}_{i}\times(\vec{S}_{1}+\vec{S}_{2}%
+\vec{S}_{3})\\
&  =J_{\parallel}s\hspace{3pt}\hat{e}_{z}\times\vec{S}_{\mathrm{tot}}%
\end{align}
or
\[
0=\sum_{j\in<i>}S_{j}^{x}=\sum_{j\in<i>}S_{j}^{y}.
\]
In modes (\ref{eq:FMeigenstates2})-(\ref{eq:FMeigenstates3}) the excitation is
equally distributed over the lattice, so that the in-plane components
$S_{1}^{\Vert}=S_{2}^{\Vert}=S_{3}^{\Vert}$. The only solution is then given
by a relative phase shift of $\frac{2\pi}{3}$ which agrees with the eigenmodes
at Dirac points $K$,$K^{\prime}$ obtained by diagonalizing the magnon Hamiltonian.

\end{document}